\documentclass[usenatbib,conference]{basii}                                       %
\usepackage[british]{babel}
\usepackage[varg]{txfonts}
\usepackage{rotating}
\usepackage{dcolumn}

\renewcommand{\vec}[1]{\mbox{\boldmath $#1$}}

\begin{document}
\title[Solar differential rotation]{Solar differential rotation: origin, models, and implications for dynamo}
\author[L.\,L.~Kitchatinov]
       {L.\,L.~Kitchatinov$^{1,2}$\thanks{Email: \texttt{kit@iszf.irk.ru}}\\
       $^1$Institute for Solar-Terrestrial Physics, P.O. Box 291, Irkutsk 664033\\
       $^2$Pulkovo Astronomical Observatory, St. Petersburg 196140, Russian Federation}

\pubyear{2011}
\volume{00}
\pagerange{\pageref{firstpage}--\pageref{lastpage}}

\date{Received \today}

\maketitle \label{firstpage}
\begin{abstract}
Helioseismology shows that the regions occupied by convection and
differential rotation inside the sun almost coincide. This supports
the leading theoretical concept for the origin of differential
rotation as a result of interaction between convection and rotation.
This talk outlines the current state of the differential rotation
theory. Numerical models based on the theory reproduce the observed
solar rotation quite closely. The models also compute meridional
flow and predict that the flow at the bottom of the convection zone is
not small compared to the surface. Theoretical predictions for
stellar differential rotation as a function of the stellar mass and
rotation rate are discussed and compared with observations.
The implications of the differential rotation models for solar and
stellar dynamos are briefly discussed.
\end{abstract}


\section{Introduction}\label{intro}
The theory of global flows on the Sun is more fortunate in getting
guidance from helioseismology than the twin-theory of global solar
magnetic fields. Though the main concepts of the differential
rotation theory were formulated before the emergence of
helioseismology, the detailed helioseismological picture of the
internal solar rotation helped to avoid spending time and efforts on
studying theories not compatible with the picture. Now, about 150
years after the discovery of differential rotation of the sun by
\citet{C1863}, the origin of differential rotation seems to be well
understood theoretically. Numerical models based on the theory
reproduce solar rotation quite closely \citep{TT04} and predictions
for the differential rotation of stars are to some extent confirmed
by observations.

This talk discusses the main processes thought of as responsible for the
formation of solar differential rotation. Then, we shall see
what the numerical models based on this theory produce for the
sun and what they predict for solar-type stars. Implications of
the results for solar and stellar dynamos are briefly discussed. The
talk is mainly focused on the mean-field theory of differential
rotation. The 3D numerical experiments were recently reviewed by
\citet{BR09}.
\section{Theory}
\subsection{Differential rotation}
Helioseismology shows that the decrease of angular velocity from the
equator to poles seen on the surface of the sun survives throughout
the convection zone up to its base but disappears shortly beneath
the base \citep{Wea97,Sea98}. The regions inside the sun occupied by
differential rotation and convection almost coincide. This supports
the theoretical concept that explains the differential rotation by
the interaction between convection and rotation. Convective motions
in rotating fluid are disturbed by the Coriolis force. The back
reaction disturbs rotation making it not uniform.

\begin{figure}
    \centerline{
    \includegraphics[width=0.5\textwidth]{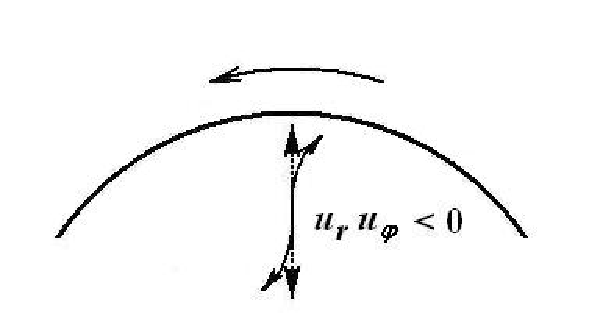}
    \includegraphics[width=0.5\textwidth]{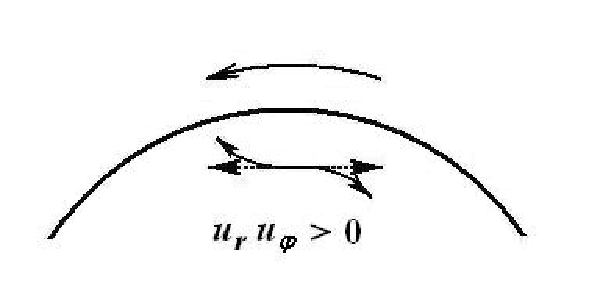}
    }
    \caption{Illustration of angular momentum transport by turbulent
        mixing. Direction of rotation is
        shown on the top. {\sl Left:} The original radial motion (dashed
        arrows) is disturbed by the Coriolis force so that the
        product $u_ru_\phi$ is negative and the angular momentum is
        transported downward. {\sl Right:} $u_ru_\phi > 0$ for
        horizontal mixing producing upward transport of the angular
        momentum. Anisotropic mixing with different intensities of
        radial and horizontal motions is required for the net flux
        of the angular momentum to emerge.
        }
    \label{f1}
\end{figure}

Considering details of the process, it is easy to see that
convective mixing along the radius tends to produce a state of
sub-rotation with angular velocity increasing with depth.
Figure~\ref{f1} shows that the fluid particles originally moving
along the radius are deflected by the Coriolis force to attain
azimuthal velocities. The product $u_ru_\phi$ is negative for both
the upward and downward original motion; $\vec{u}$ is convective
velocity, the standard spherical coordinates $(r,\theta,\phi )$ are
used. The negative value, $u_ru_\phi < 0$, means that the angular
momentum is transported downward. This effect is so clear that
almost every year brings new papers stating that it should be a
general rule that angular velocity in mixed spherical bodies
increases with depth.

The left panel of Fig.~\ref{f1} does not show a complete picture,
however. The right-hand panel shows that horizontal mixing tends to produce
the state of super-rotation by transporting angular momentum upward.
The cross-component, $Q_{r\phi}$, of the correlation tensor $Q_{ij}
= \langle u_iu_j\rangle$ can be estimated as follows,
\begin{equation}
    Q^\Lambda_{r\phi} = 2\tau\Omega\
    \left(\langle u_\phi^2\rangle - \langle u_r^2\rangle \right)
    \ \sin\theta ,
    \label{1}
\end{equation}
where $\Omega$ is angular velocity, and $\tau$ is convective
turnover time, the meaning of the upper index $\Lambda$ will be
explained shortly. In order for the net (convective) flux of angular
momentum to exist, intensities of radial and horizontal convective
mixing should differ and the direction of the flux is defined by the
sense of the anisotropy of the mixing \citep{L41,W46,B51}.

The right part of Eq.~(\ref{1}) involves a key parameter of the
differential rotation theory,
\begin{equation}
    \Omega^* = 2\tau\Omega,
    \label{2}
\end{equation}
named the Coriolis number. The parameter measures the intensity of
interaction between convection and rotation. The dramatic complication
of the theory comes from the fact that the sun and absolute majority
of cool stars have $\Omega^* > 1$ in the bulk of their convection
zones. This means that the linear estimation (\ref{1}) does not
apply and the theory should be nonlinear in $\Omega^*$. Nonlinear
derivations of angular momentum fluxes show that for the nearly
adiabatic stratification of convection zones the fluxes are parallel
to the rotation axis and point to the equatorial plane in the rapid
rotation case, $\Omega^* \gg 1$ \citep{KR93}. At small depths in
the solar convection zone, the Coriolis number is smaller than one.
The angular momentum fluxes have a radial inward direction in this
case \citep{Kea04,KR05}. As $\Omega^*$ increases with increasing
depth in the convection zone, the fluxes change from an inward radial
direction to an equatorward one parallel to the rotation axis.

The ability of convection to transport angular momentum even in the case
of {\em rigid} rotation was named the $\Lambda$-effect \citep{R89}.
The rigidity of rotation is emphasized because turbulent viscosity can
also transport angular momentum if rotation is not uniform. The
$Q^\Lambda$ of Eq.~(\ref{1}) is only a part of the total correlation of
convective velocities, namely the part representing the
$\Lambda$-effect. The total correlation includes the viscous part,
$Q^\nu$, also:
\begin{equation}
    Q_{ij} = Q^\Lambda_{ij} + Q^\nu_{ij},\ \ \ Q^\nu_{ij} = - {\cal
    N}_{ijkl}\frac{\partial V_k}{\partial r_l},
    \label{3}
\end{equation}
where ${\vec V}$ is the large-scale velocity and ${\cal N}_{ijkl}$
is the turbulent viscosity tensor. Viscous fluxes of angular
momentum tend to reduce differential rotation. They increase with
the inhomogeneity of angular velocity. Non-diffusive fluxes (the
$\Lambda$-effect) depend on the angular velocity, not on its
gradient. A steady state of differential rotation can \lq to the
first approximation' be understood as a balance between the
$\Lambda$-effect and the eddy viscosity.

The approximation is, however, rather rough because it misses the
important contribution to angular momentum transport made by the
global meridional flow. The steady mean-field equation for the
angular momentum balance reads \citep{K05}
\begin{equation}
    \mathrm{div}\left( \rho r \sin\theta\langle u_\phi\vec{u}\rangle
    + \rho r^2\sin^2\theta\ \Omega\vec{V}^\mathrm{m}\right) = 0 ,
    \label{4}
\end{equation}
where $\vec{V}^\mathrm{m}$ is the meridional flow velocity.
Substitution of an explicit expression for the correlations
(\ref{3}) of convective velocities into Eq.\,(\ref{4}) gives an
equation for the angular velocity. The equation is, however, not
closed because it includes the yet undefinite meridional flow. The
flow cannot be neglected or prescribed because meridional flow
is produced by differential rotation \citep{K63}. Global flow in
the convection zone of a star represents a self-regulating system:
differential rotation produces meridional flow which in turn
modifies the differential rotation.
\subsection{Meridional flow}\label{mery}
The origin of meridional flow is well illustrated by the
(steady) equation for this flow,
\begin{equation}
    {\cal D}(\vec{V}^\mathrm{m})\ =\ \sin\theta\ r{\partial\Omega^2\over\partial z}\
    -\ {g\over c_{\rm p} r}{\partial S\over\partial\theta},
    \label{5}
\end{equation}
where $z = r\cos\theta$ is the distance from the equatorial plane,
$S$ is the specific entropy, $g$ is gravity, and $c_\mathrm{p}$
is the specific heat at constant pressure. The left part of this
equation describes the viscous drag on the meridional flow (we will not
need a rather bulky explicit expression for this term in the
discussion to follow). The right part includes the two sources of the
meridional flow.

The first term on the right of Eq.\,(\ref{5}) represents centrifugal
driving. If the angular velocity varies with $z$ to decrease with
distance to the equatorial plane, as it does in the sun, the
centrifugal force produces a torque driving a flow to the poles near
the surface and a flow to the equator near the bottom of the
convection zone.

The second term on the right of Eq.\,(\ref{5}) involves the
so-called \lq baroclinic driving' known also as the source of the
\lq thermal wind'. If polar regions are warmer then the equator,
as they are on the sun \citep{Rea08}, the baroclinic driving
counteracts the centrifugal driving.

For solar and stellar conditions, each of the two terms on the right
side of Eq.\,(\ref{5}) is large compared to the left side.
Therefore, the two terms nearly balance each other - the condition
known as the Taylor-Proudman balance. The balance is maintained
mainly via the influence of meridional flow on the rotation law
\citep{D89}.

Equations (\ref{4}) and (\ref{5}) again do not represent a
closed system because the latitudinal entropy gradient is not
defined.
\subsection{Differential temperature}
Turbulent heat transport in rotating convection zones is
anisotropic. Not only does the heat transport coefficient depend on
latitude \citep{W65}, but the direction of the convective heat flux is
not aligned with the entropy gradient \citep{Rea05}.

The convective heat flux,
\begin{equation}
    F^\mathrm{conv}_i = -\rho T \chi_{ij}\frac{\partial S}{\partial
    r_j} ,
    \label{6}
\end{equation}
depends on the structure of the thermal conductivity tensor,
\begin{eqnarray}
    \chi_{ij} &=& \chi_{_\mathrm{T}}\left( \phi(\Omega^*)\delta_{ij} +
    \phi_\|(\Omega^*)\hat\Omega_i\hat\Omega_j\right),
    \label{7}
    \\
    \chi_{_\mathrm{T}} &=& -\frac{\tau\ell^2g}{12
    c_\mathrm{p}}\frac{\partial S}{\partial r},
    \label{8}
\end{eqnarray}
where $\hat{\vec\Omega} = {\vec\Omega}/\Omega$ is unit vector along
the rotation axis and the functions $\phi(\Omega^*)$,
$\phi_\|(\Omega^*)$ of the Coriolis number (\ref{2}) involve the
rotationally induced anisotropy and quenching of the diffusivity
\citep{Kea94}. Even if entropy varies mainly with radius, the heat
flux (\ref{6}) deviates from the radial direction towards the poles.
The poleward latitudinal heat flux is the main reason for the \lq
differential temperature' phenomenon in the mean-field theory. There
were multiple attempts to measure differential temperature on the
sun. Recent observations by \citet{Rea08} suggest that the solar
poles are warmer than the equator by about 2.5\,K.

This small temperature  difference between the equator and poles on
the very hot sun has important hydrodynamical consequences. It can
lead to differential rotation even without the $\Lambda$-effect
\citep{DR71}: the differential temperature produces meridional flow
by the baroclinic driving of Eq.\,(\ref{5}) and the flow in turn
transports angular momentum (\ref{4}) to produce differential
rotation. Models based on the anisotropic heat transport, however,
do not reproduce the solar rotation. Nevertheless, only when
differential temperature is accounted for can the
helioseismologically detected internal rotation of the sun be
reproduced \citep{KR95,Mea06}.

The angular momentum equation (\ref{4}) together with the meridional
flow equation (\ref{5}) and entropy equation with the heat flux
(\ref{6}) represent a closed system that can be solved numerically
for the distributions of angular velocity, meridional flow and
entropy in a stellar convection zone. It should be noted that all
that is needed for the numerical modeling - the $\Lambda$-effect,
thermal conductivities and eddy viscosities - have been derived
within the same approach. As a result, uncertainty in the model
design was minimized and the mean-field models practically do not
involve free parameters.
\section{Models}
\subsection{The Sun}
\begin{figure}
    \centerline{
    \includegraphics[width=0.85\textwidth]{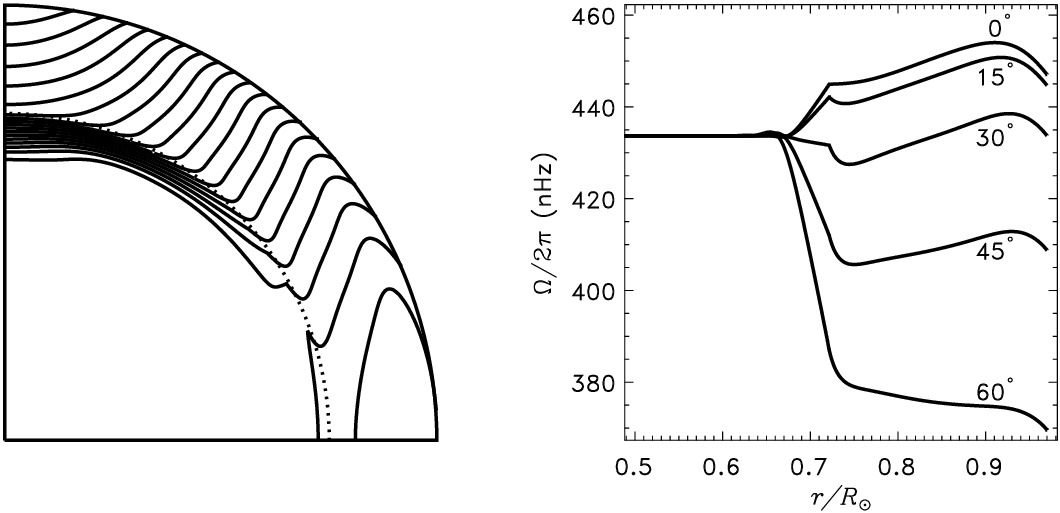}
    }
    \caption{Angular velocity isolines ({\sl left}) and depth
    profiles of the rotation rate for several latitudes  ({\sl
    right}) after the mean-field model of \citet{KO11}.
        }
    \label{f2}
\end{figure}

Figure~\ref{f2} shows the internal solar rotation computed with the
mean-field model. The results are similar to the helioseismological
rotation law.

The discussion in the preceding section refers to the convection zone only.
Figure~\ref{f2} includes the tachocline region and the deeper
radiation zone. The Figure was produced by joint use of two
different models. Physical conditions in convection and radiation
zones differ so much that it is not possible to cover both in one
model. The rotation of the radiation core of Fig.~\ref{f2} was computed
with the magnetic tachocline model of \citet{RK97}. The modeling of
the tachocline does not influence the computation of the differential
rotation of the convection zone in any way but just uses the results of
this computation as a boundary condition. There is no space for
discussing tachocline physics here. We just mention that up to
now it has been possible to explain simultaneously the uniform rotation
in the deep radiation core and a slender tachocline on its top only
by an effect of an internal relic magnetic field
\citep{CM93,RK97,MC99,D10}. The tachocline of Fig~\ref{f2} is
produced by a weak internal poloidal field of about $10^{-2}$~Gauss.

The angular velocity distributions in theoretical models are
symmetric about the equator and regular near the poles. This implies
that the angular velocity isolines are normal to both the equatorial
plane and the rotation axis. The isorotational surfaces,
cylinder-shaped near the equator and disk-shaped near the poles,
are, therefore, elementary consequences of the global symmetry of
the problem (for alternative opinion see
\citeauthor{B09}~\citeyear{B09}). The nearly radial isolines of
Fig.~\ref{f2} at middle latitudes are, however, not trivial. They
are the consequence of the Taylor-Proudman balance in the bulk of
the convection zone (see Section \ref{mery} above). Only with
allowance for the effect of differential temperature can deviations
from cylinder-shaped isorotation surfaces be reproduced.

\begin{figure}
    \centerline{
    \includegraphics[width=0.85\textwidth]{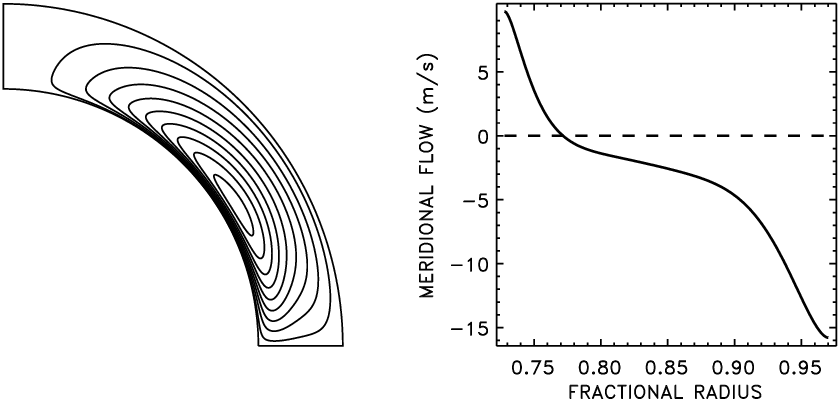}
    }
    \caption{Meridional flow after the same model as Fig.~\ref{f2}. {\sl
    Left}: Meridional flow stream-lines. {\sl Right:} Radial profile
    of meridional velocity for $45^\circ$ latitude. Negative velocity
    means poleward flow.
        }
    \label{f3}
\end{figure}

Differential rotation models compute also the distributions of
entropy and the meridional flow. The flow is increasingly recognized
as important for the solar dynamo \citep{C08}. The flow at small
depths can be probed by helioseismology \citep{ZK04}. Theoretical
modeling remains, however, the only source of knowledge on the deep
meridional flow. The typical structure of the simulated circulation
is shown in Fig.~\ref{f3}. The plot shows the flow distribution up
to the base of the convection zone. Beneath the base, the meridional
velocity rapidly decreases with depth \citep{GM04}.

Meridional flow results from a (small) disbalance between two
terms in the right side of Eq.~(\ref{5}). Observations of the flow
\citep{Kea93} indicate that some deviations from the Taylor-Proudman
balance are present in the sun. The flow of Fig.~\ref{f3} is
relatively small in the bulk of the convection zone and increases towards the
zone boundaries. Such a structure of the flow is related to the
boundary layers \citep{D89}. The Taylor-Proudman balance is not
compatible with stress-free boundary conditions. As a result,
boundary layers form where the balance is violated and the sources
of meridional flow are relatively large. The bottom flow of
Fig.~\ref{f3} is faster than usually assumed in advection-dominated
dynamo models. The first dynamo-model with a fast near-bottom flow
was recently produced by \citet{PK11}.

Another implication for dynamo models is related to the value of the
magnetic eddy diffusivity. The diffusivities in the differential
rotation models are not prescribed but expressed in terms of the
entropy gradient, as it is done in Eq.~(\ref{8}) for turbulent
thermal diffusion. Characteristic values of the resulting turbulent
viscosities and diffusivities are about $10^{13}$\,cm$^2$s$^{-1}$
for the sun; global circulation models with smaller diffusivities
are unstable \citep{Tea94}. Theories of turbulent transport
coefficients and 3D simulations \citep{Yea03} both suggest that the
magnetic Prandtl number is of the order of one, i.e., the eddy
magnetic diffusivity is also about $10^{13}$\,cm$^2$s$^{-1}$. This
value is larger than usually assumed in solar dynamo models.
\subsection{Cool and Solar-Type Stars}
The mean-field models for differential rotation can, of course, be
applied to convective stars other than the sun. The models, which do
not include the effects of magnetic fields, always predict solar-type
rotation with the equator rotating faster than the poles. They also predict
that dependence on rotation rate for a star of given structure is
relatively small. The main prediction, however, is that the surface
differential rotation increases with stellar mass \citep{KR99}.

\begin{figure}
    \centerline{
    \includegraphics[width=0.47\textwidth,height=0.363\textwidth]{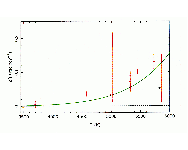}
    \hspace{0.2truecm}
    \includegraphics[width=0.49\textwidth]{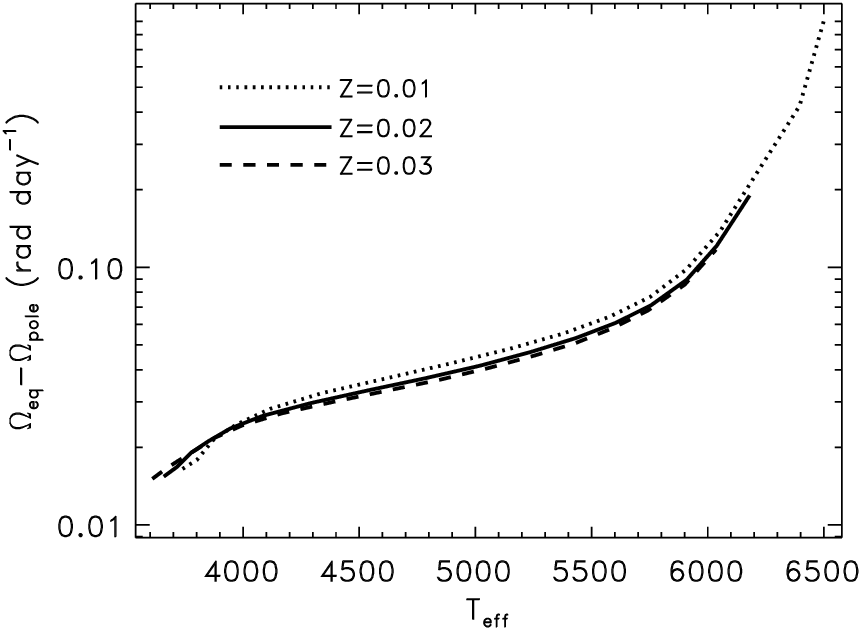}
    }
    \caption{{\sl Left:} Dependence of surface differential rotation
    on effective temperature for rapidly rotating solar analogues
    observed using the Doppler imaging techniques
    (\citeauthor{Bea05}\citeyear{Bea05}; figure courtesy of John
    Barnes). {\sl Right:} Same dependence computed with the
    differential rotation model \citep{KO11}. Lines of different style
    correspond to different metallicities $Z$.
        }
    \label{f4}
\end{figure}

The predictions are in at least qualitative agreement with
observations. Differential rotation of two solar twins rotating
about three times faster than the sun was measured recently using
high precision photometry of the {\sl MOST}-mission
\citep{Cea06,Wea07}. In both cases, the amount of the surface
differential rotation was close to solar value. Differential
rotation measurements by Doppler imaging for young rapidly rotating
stars were summarized by \citet{Bea05}. Figure~\ref{f4} compares the
dependence on stellar surface temperature they found with
computations of \citet{KO11}.

\begin{figure}
    \begin{center}
    \begin{tabular}{p{6.5cm}cp{4.2cm}}
    \raisebox{-\height}{\includegraphics[width=0.5\textwidth]{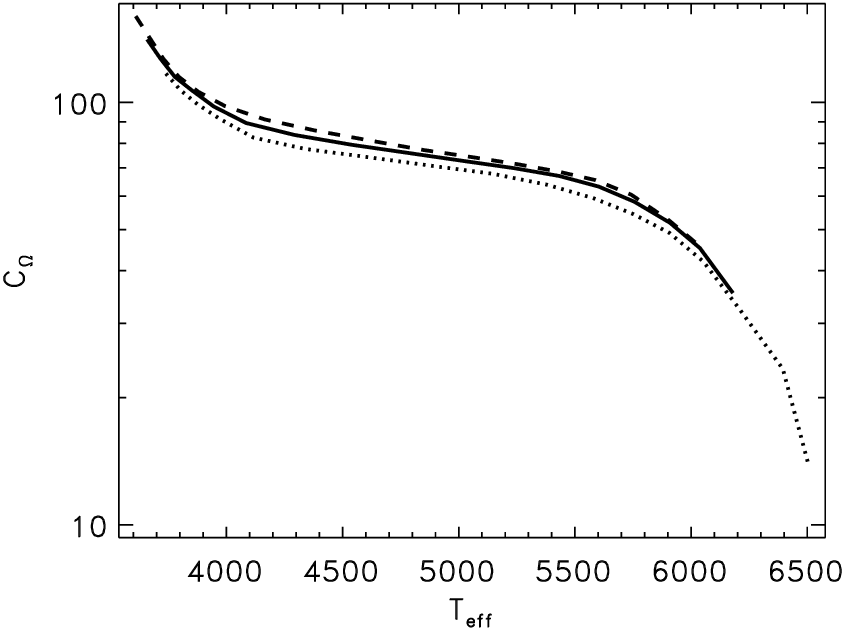}} & \quad &
    \caption{$C_\Omega$ dynamo number (\ref{9}) as the function of stellar
        surface temperature after the same model as the right panel of
        Fig.~\ref{f4}. Different line styles show the results for
        different chemical compositions.
        }
    \end{tabular}
    \end{center}
    \label{f5}
\end{figure}

Both plots of Fig.~\ref{4} suggest that the hottest convective stars
possess the largest differential rotation. As the rotational shear is
important for dynamos, the question arises whether the strong
differential rotation of F-stars implies over-normal dynamo
activity? Surprisingly, the answer is negative. Dynamo theory
estimates the efficiency of differential rotation in generating
magnetic fields by the variety of the magnetic Reynolds number
conventionally notated as $C_\Omega$,
\begin{equation}
    C_\Omega = \frac{\Delta\Omega H^2}{\eta_{_\mathrm{T}}} ,
    \label{9}
\end{equation}
where $\Delta\Omega$ is the angular velocity variation within the
convection zone, $H$ is the convection zone thickness, and
$\eta_{_\mathrm{T}}$ is the turbulent magnetic diffusivity. Figure~5
shows the dependence of $C_\Omega$ on stellar effective temperature
after the same model as the differential rotation plot of
Fig.~\ref{f4}. The two Figures display, however, opposite trends.
The $C_\Omega$-parameter decreases with temperature. The large
differential rotation of F-stars is much less efficient at producing
toroidal magnetic fields than the almost uniform rotation of
M-stars. This is in agreement with the old idea of \citet{DL78} that
convective dynamos cease to operate at about spectral type F6.
\section*{Acknowledgements}
The author is thankful for the support by the Russian Foundation for
Basic Research (Projects 11-02-08020-z, 10-02-00148-a).

\newpage

\label{lastpage}
\end{document}